\documentclass{kapproc}

\usepackage{procps}
\usepackage{amstext,amsgen,latexsym}
\usepackage{amstext,amssymb,amsfonts,latexsym}

\makeatletter
\renewcommand{\normalsize}{%
    \@setfontsize\normalsize\@xpt\@xiipt
    \abovedisplayskip 10\p@ \@plus2\p@ \@minus5\p@
    \abovedisplayshortskip \z@ \@plus3\p@
    \belowdisplayshortskip 6\p@ \@plus3\p@ \@minus3\p@
    \belowdisplayskip \abovedisplayskip
    \let\@listi\@listI}
\normalsize
\makeatother

\normallatexbib

\let\footnote\savefootnote
\let\footnotetext\savefootnotetext

 \newcommand{\bs}{\bigskip}
 
 \newcommand{\n}{\noindent}
 
 \newcommand{\hs}[1]{\hspace*{ #1 mm}}
 \newcommand{\vs}[1]{\vspace*{ #1 mm}}

 \newcommand{\setempty}{\mathrm{\O}}

 \newcommand{\nat}{\mathbb{N}}

 \newcommand{\complex}{\mathbb{C}}
 \newcommand{\appcomplex}{\tilde{\mathbb{C}}}

 \newcommand{\prob}{{\mathrm{Prob}}}

 \newcommand{\co}{\mathrm{co}\mbox{-}}

 \newcommand{\eg}{\textrm{e.g.},\hspace*{2mm}}
 
 \newcommand{\etalc}{\textrm{et al.}}

 \newcommand{\CC}{{\cal C}}
 
 \newcommand{\DD}{{\cal D}}
 \newcommand{\HH}{{\cal H}}

 \newcommand{\UU}{{\cal U}}

 \newcommand{\p}{\mathrm{P}}
 \newcommand{\np}{\mathrm{NP}}
 \newcommand{\ph}{\mathrm{PH}}

 \newcommand{\ma}{\mathrm{MA}}

 \newcommand{\up}{\mathrm{UP}}

 \newcommand{\dexp}{\mathrm{EXP}}
 \newcommand{\nexp}{\mathrm{NEXP}}

 \newcommand{\cequalp}{\mathrm{C}_{=}\mathrm{P}}

 \newcommand{\nqp}{\mathrm{NQP}}
 \newcommand{\bqp}{\mathrm{BQP}}
 \newcommand{\pqp}{\mathrm{PQP}}

 \newcommand{\qma}{\mathrm{QMA}}

 \newcommand{\sharpqp}{\#\mathrm{QP}}

 \newcommand{\qoptsharpqp}{\mathrm{Qopt}\#\mathrm{QP}}
 \newcommand{\qoptsharpsigmaqh}[1]%
     {\mathrm{Qopt}\#\Sigma^{\mathrm{QP}}_{ #1 }}
 \newcommand{\qoptsharppiqh}[1]%
     {\mathrm{Qopt}\#\Pi^{\mathrm{QP}}_{ #1 }}

 \newcommand{\bp}{\mathrm{BP}\cdot}
 
 \newcommand{\qe}{\exists^{\mathrm{Q}}\cdot}
 \newcommand{\qa}{\forall^{\mathrm{Q}}\cdot}

 \newcommand{\sigmah}[1]{\Sigma^{\mathrm{P}}_{#1}}
 \newcommand{\pih}[1]{\Pi^{\mathrm{P}}_{#1}}

 \newcommand{\sigmaqh}[1]{\Sigma^{\mathrm{QP}}_{#1}}
 \newcommand{\piqh}[1]{\Pi^{\mathrm{QP}}_{#1}}
 
 \newcommand{\qph}{\mathrm{QPH}}

 \newcommand{\deltaexph}[1]{\Delta^{\mathrm{EXP}}_{#1}}
 \newcommand{\sigmaexph}[1]{\Sigma^{\mathrm{EXP}}_{#1}}
 \newcommand{\piexph}[1]{\Pi^{\mathrm{EXP}}_{#1}}
 \newcommand{\exph}{\mathrm{EXPH}}

 \newcommand{\IFF}{\Longleftrightarrow}

 \def\bbox{\vrule height6pt width6pt depth1pt}

 \newtheorem{lemma}[theorem]{Lemma}
 \newtheorem{proposition}[theorem]{Proposition}

 \newtheorem{definition}[theorem]{Definition}

 \newenvironment{proof}{\par \noindent
            {\bf Proof. \hs{1}}}{\hfill$\Box$ \vspace*{3mm}}

 \newenvironment{proofsketch}{\par \noindent
            {\bf Proof Sketch. \hs{1}}}{\hfill\bbox \vspace*{3mm}}

 \newenvironment{proofof}[1]{\vspace*{5mm} \par \noindent
         {\bf Proof of #1.\hs{1}}}{\hfill$\Box$ \vspace*{3mm}}

 \newcommand{\ceilings}[1]{\lceil #1 \rceil}
 
 \newcommand{\pair}[1]{\langle #1 \rangle}

 \newcommand{\dtime}[1]{{\mathrm{DTIME}}(#1)}

 \newcommand{\qubit}[1]{| #1 \rangle}
 \newcommand{\bra}[1]{\langle #1 |}
 \newcommand{\ket}[1]{| #1 \rangle}
 \newcommand{\measure}[2]{\langle #1 | #2 \rangle}
 \newcommand{\qustrings}{\Phi}
 
 \newcommand{\ignore}[1]{}
 \newcommand{\quantset}[1]{{}^* #1}

\begin{document}

\articletitle[Quantum NP and a Quantum Hierarchy]{Quantum NP 
and a Quantum Hierarchy}
\articlesubtitle{(Extended Abstract)}
\author{Tomoyuki Yamakami}
\affil{School of Information Technology and Engineering \\
 University of Ottawa, Ottawa, Ontario, Canada  K1N 6N5}
\email{yamakami@site.uottawa.ca}

\begin{abstract}
The complexity class $\np$ is quintessential and ubiquitous in
theoretical computer science.  Two different approaches have been made
to define ``Quantum NP,'' the quantum analogue of $\np$: $\nqp$ by
Adleman, DeMarrais, and Huang, and $\qma$ by Knill, Kitaev, and
Watrous. {}From an operator point of view, $\np$ can be viewed as the
result of the $\exists$-operator applied to $\p$.  Recently, Green,
Homer, Moore, and Pollett proposed its quantum version, called the
N-operator, which is an abstraction of $\nqp$. This paper introduces
the $\exists^{\mathrm{Q}}$-operator, which is an abstraction of
$\qma$, and its complement, the $\forall^{\mathrm{Q}}$-operator. These
operators not only define Quantum NP but also build a quantum
hierarchy, similar to the Meyer-Stockmeyer polynomial hierarchy, based
on two-sided bounded-error quantum computation.
\end{abstract}

\bs\bs

\section{What is Quantum NP?}\label{sec:quantum-NP}

\n Computational complexity theory based on a Turing 
machine (TM, for short) was formulated in the 1960s.  The complexity
class $\np$ was later introduced as the collection of sets that are
recognized by nondeterministic TMs in polynomial time. By the earlier
work of Cook, Levin, and Karp, $\np$ was quickly identified as a
central notion in complexity theory by means of
$\np$-completeness. $\np$ has since then exhibited its rich structure
and is proven to be vital to many fields of theoretical computer
science.  Meyer and Stockmeyer \cite{MS72} further extended $\np$ into
a hierarchy, known as the {\em polynomial (time) hierarchy}. This
hierarchy has inspired many tools and techniques, e.g., circuit
lower-bound proofs and micro hierarchies within $\np$. There is known
to be a relativized world where the hierarchy forms an infinite
hierarchy.  It is thus natural to consider a quantum analogue of
$\np$, dubbed as ``Quantum NP,'' and its extension. Several approaches
have been made over the years to define Quantum NP.

As is known, $\np$ can be characterized in several different
manners. As the first example, $\np$ can be characterized by
probabilistic TMs with positive acceptance probability.  Adleman
\etalc~\cite{ADH97} introduced the complexity class
$\nqp$ as a quantum extension of this probabilistic
characterization. Subsequently, $\nqp$ (even with arbitrary complex
amplitudes) was shown to coincide with the classical counting class
$\co\cequalp$
\cite{FR99,FGHP99,YY99}. This shows the power of quantum computation.

$\np$ can be also characterized by logical quantifiers over classical
(binary) strings of polynomial length. This is also known as the
``guess-and-check'' process. Knill \cite{Kni96}, Kitaev
\cite{Kit99}, and Watrous \cite{Wat00} studied the complexity class
$\qma$ (named by Watrous), which can be viewed as a quantum extension
of the aforementioned quantifier characterization of $\np$. In their
definition, a quantifier bounds a quantum state instead of a classical
string. We call such a quantifier a {\em quantum quantifier} to
emphasize the scope of the quantifier being quantum states. Using this
terminology, any set in $\qma$ is defined with the use of a single
quantum quantifier over polynomial-size quantum states. It appears
that a quantum quantifier behaves in quite a distinctive manner. For
instance, Kobayashi
\etalc~\cite{KMY01} recently pointed out that allowing multiple
quantum quantifiers may increase the complexity of $\qma$ due to
quantum entanglement (in \cite{KMY01}, $\qma(k)$ is defined with $k$
quantum quantifiers).

{}From a different aspect, we can view the process of defining $\np$
as an application of 
an operator that transforms a class $\CC$ to another class $\DD$.
For example, we write $\co\CC$ to denote the class $\{A\mid
\overline{A}\in\CC\}$, where $\overline{A}$ is the complement of
$A$. This prefix ``co'' in $\co\CC$ can be considered as the {\em
complementation operator} that builds $\co\CC$ from $\CC$.  Other
examples are Sch{\"o}ning's BP-operator \cite{Sch89} and Wagner's
C-operator \cite{Wag86}. The classical existential quantifier
naturally induces the so-called {\em $\exists$-operator}. With this
$\exists$-operator, $\np$ is defined as $\exists\cdot\p$.  Similarly,
we can consider a quantum analogue of the $\exists$-operator. One
possible analogue was recently proposed by Green
\etalc~\cite{GHMP02}. They introduced the N-operator, which is an
abstraction of $\nqp$.

To make the most of quantum nature, we define in this paper a quantum
operator that expands the quantum existential quantifier used for
$\qma$ and $\qma(k)$. This quantum operator is called the {\em
$\exists^{\mathrm{Q}}$-operator} (whose complement is the {\em
$\forall^{\mathrm{Q}}$-operators}). These quantum operators give a new
definition for Quantum NP and its expansion, a quantum analogue of the
polynomial hierarchy. Our quantum operators, however, require a more
general framework than the existing one. In the subsequent section, we
discuss a general framework for the quantum operators.

\section{Toward a General Framework for Quantum Operators}

\n Let our alphabet
$\Sigma$ be $\{0,1\}$ throughout this paper. Let $\nat$ be the set of
all nonnegative integers and set $\nat^{+}=\nat-\{0\}$.  To describe a
quantum state, we use Dirac's ket notation $\qubit{\phi}$.  Write
$\HH_n$ to denote a Hilbert space of dimension $n$. In comparison with
a classical (binary) string, we use the terminology, a {\em quantum
string} ({\em qustring}, for short) {\em of size $n$}, to mean a
unit-norm vector in $\HH_{2^n}$. For such a qustring $\qubit{\phi}$,
$\ell(\qubit{\phi})$ denotes the {\em size} of $\qubit{\phi}$. We use
the notation $\qustrings_n$ for each $n\in\nat$ to denote the
collection of all qustrings of size $n$ and thus,
$\qustrings_n\subseteq\HH_{2^n}$.  Let
$\qustrings_{\infty}=\bigcup_{n\geq0}\qustrings_n$, the set of all
finite-size qustrings.

We use a multi-tape quantum Turing machine (QTM), defined in
\cite{BV97,Yam99a}, as a mathematical model of quantum computations. A
multi-tape QTM is equipped with two-way infinite tapes, tape heads,
and a finite-control unit.  We assume in this paper the following
technical restriction on each QTM: a QTM is always designed so that
all computation paths on each input terminate at the same time by
entering its unique halting state after writing $0$ (rejection) or $1$
(acceptance) in the start cell of the designated output tape (see
\cite{Yam99a} for the discussion on the {\em timing problem}). Thus,
the length of a computation path on input $x$ is regarded as the {\em
running time} of the QTM on $x$. The transition function $\delta$ of a
QTM can be seen as an operator (called a {\em time-evolution
operator}) that transforms a superposition of configurations at time
$t$ to another superposition of configurations at time $t+1$.  A QTM
is called {\em well-formed} if its time-evolution operator is
unitary. Moreover, a QTM is said to have {\em
$\appcomplex$-amplitudes} if all amplitudes in $\delta$ are drawn from
set $\appcomplex$, where $\appcomplex$ is the set of all complex
numbers whose real and imaginary parts are approximated
deterministically to within $2^{-n}$ in time polynomial in $n$.  For a
well-formed QTM $M$ and an input $\qubit{\phi}$, the notation
$\prob_{M}[M(\qubit{\phi})=1]$ denotes the acceptance probability of
$M$ on input $\qubit{\phi}$. Similarly, $\prob_{M}[M(\qubit{\phi})=0]$
denotes the rejection probability of $M$ on $\qubit{\phi}$.

\subsection{{}From Classical Inputs to Quantum Inputs}

\n We have used classical (binary) strings as standard 
inputs given into quantum computations. As a result, any quantum
complexity class, such as $\nqp$ or $\bqp$ \cite{BV97}, is defined to
be a collection of subsets of $\Sigma^*$. Since a QTM acts as a
unitary operator, it is legitimate to feed the QTM with a quantum
state as an input. We call such an input a {\em quantum input} for
clarity. As in the definition of $\qma(k)$, for instance, such quantum
inputs play an essential role.  We thus need to expand a set of
strings to a set of qustrings by considering a qustring as an input
given to an underlying QTM. We use the following notation. For each
$m,n\in\nat^{+}$, let $\qustrings_{n}^m$ denote the collection of all
$m$-tuples $(\qubit{\phi_1},\qubit{\phi_2},\ldots,\qubit{\phi_m})$
such that each $\qubit{\phi_i}$ is a qustring of size $n$. Such an
$m$-tuple is expressed as $\qubit{\vec{\phi}}$ and also seen as a
tensor product $\qubit{\phi_1}\qubit{\phi_2}\cdots \qubit{\phi_m}$
when the size of each $\qubit{\phi_i}$ is known. For brevity, the
notation $\ell(\qubit{\vec{\phi}})$ means the sum
$\sum_{i=1}^{m}\ell(\qubit{\phi_i})$. We also set
$\qustrings_{\infty}^{m}=
\bigcup_{n\geq1}\qustrings_{n}^{m}$ and
$\qustrings_{\infty}^{*}=
\bigcup_{m\geq1}\qustrings_{\infty}^m$.

The introduction of quantum inputs gives rise to an important issue,
which is not present in the classical framework: the duplication of an
input. The repetition of a quantum computation on a classical input is
seen in, e.g., the proof of $\bqp^{\bqp}=\bqp$
\cite{BBBV97}. Nevertheless, 
the situation may change when we deal with a quantum input. Since a
fundamental principle of quantum computation, the so-called {\em
no-cloning theorem}, interdicts the duplication of an arbitrary
quantum input, we cannot redo even the same quantum computation on a
single quantum input unless the copies of the quantum input are given
{\em a priori}. To establish a coherent but concise theory of quantum
computation over $\qustrings_{\infty}^{*}$, we need to allow the
underlying quantum computation to access the quantum input repeatedly
without disturbing other quantum states. Schematically, we supply a
sufficient number of its copies as ``auxiliary inputs.'' This
guarantees the quantum extension of many existing complexity classes,
such as $\bqp$, to enjoy the same structural properties.

For later convenience, we first expand the function class $\sharpqp$
\cite{Yam99b}, which originally consists of certain
quantum functions mapping from $\Sigma^*$ to the unit real interval
$[0,1]$. The notation $\quantset{\sharpqp}$ is given in this paper to
denote the corresponding extension---the collection of quantum
functions mapping from $\qustrings_{\infty}^{*}$ to $[0,1]$. Since
$\qustrings_{\infty}^{*}$ is a continuous space, these quantum
functions are inherently continuous. For simplicity, write
$\qubit{\vec{\phi}}^{\otimes k}$ for $k$ copies of
$\qubit{\vec{\phi}}$, which can be viewed as a tensor product of $k$
identical $\qubit{\vec{\phi}}$'s (as long as the size of
$\qubit{\vec{\phi}}$ is known).

\begin{definition}\label{def:sharpqp}
A function $f$ from $\qustrings_{\infty}^{*}$ to $[0,1]$ is in
$\quantset{\sharpqp}$ if there exist a polynomial $q$ and a
polynomial-time, $\appcomplex$-amplitude, well-formed QTM $M$ such
that, for every $m\in\nat^{+}$ and every
$\qubit{\vec{\phi}}\in\qustrings_{\infty}^m$, $f(\qubit{\vec{\phi}}) =
\prob_{M}[M(\qubit{\vec{\phi}}^{\otimes q(\ell(\qubit{\vec{\phi}}))})
=1]$.
\end{definition}

We reserve the standard notation $\sharpqp$ to denote the class of
quantum functions whose domains are $\Sigma^*$ (i.e., those functions
are obtained from Definition
\ref{def:sharpqp} by replacing $\qubit{\vec{\phi}}$
with $x$ from $\Sigma^*$).

To distinguish a set of qustrings from a set of classical strings, we
use the terminology, a {\em quantum set}, for a set
$A\subseteq\qustrings_{\infty}^{*}$. A collection of quantum sets is
called a {\em quantum complexity class} (which conventionally refers
to any classical class related to quantum computations). {}From a
different perspective, a classical set can be viewed as a
``projection'' of its corresponding quantum set. For a quantum set
$A\subseteq\qustrings_{\infty}^{*}$, its {\em classical part}
$\check{A}$ is given as follows:
\[
\check{A}=
\{\pair{s_1,s_2,\ldots,s_m}\mid 
m\in\nat^{+},s_1,\ldots,s_m\in\Sigma^*,
(\qubit{s_1},\qubit{s_2},\ldots,\qubit{s_m})\in A \},
\]
where $\pair{\hs{2}}$ is an appropriate {\em pairing function} from
$\bigcup_{m\geq1}(\Sigma^*)^m$ to $\Sigma^*$.  Thus, any quantum class
$\CC$ naturally induces its {\em classical part} $\{\check{A}\mid
A\in\CC\}$.  In a similar way, $\sharpqp$ is also viewed as the
``projection'' of $\quantset{\sharpqp}$.

A relativized version of $\quantset{\sharpqp}$ is defined by
substituting oracle QTMs for non-oracle QTMs in Definition
\ref{def:sharpqp}, where an {\em oracle QTM} can make a query of the
form $\qubit{x}\qubit{b}$ ($x\in\Sigma^*$ and $b\in\{0,1\}$) by which
oracle $A$ transforms $\qubit{x}\qubit{b}$ into $(-1)^{b\cdot
A(x)}\qubit{x}\qubit{b}$ in a single step.

\subsection{{}From Decision Problems to Partial Decision Problems}

\n We described in the previous subsection how to 
expand classical sets to quantum sets. The next step might be to
expand well-known complexity classes, such as $\nqp$ and $\bqp$, to
classes of quantum sets. Unfortunately, since $\qustrings_{\infty}$ is
a continuous space, we cannot expand all classical classes in this way
(for example, $\bqp$).  One of the resolutions is to consider
``partial'' decision problems. (See, \eg \cite{DK00} for classical
partial decision problems.)  In this paper, we define a {\em partial
decision problem} to be a pair $(A,B)$ such that
$A,B\subseteq\qustrings_{\infty}^{*}$ and $A\cap B=\setempty$, where
$A$ indicates a set of accepted qustrings and $B$ indicates a set of
rejected qustrings. The {\em legal region} of $(A,B)$ is $A\cup
B$. For consistency with classical decision problems, we should refer
$A$ to as $(A,\overline{A})$, where $\overline{A}
=\qustrings_{\infty}^{*}-A$, and call it a {\em total decision
problem}. The notions of {\em inclusion}, {\em union}, and {\em
complement} are introduced in the following manner: let $(A,B)$ and
$(C,D)$ be any partial decision problems and let $E$ be the
intersection of their legal regions; that is, $(A\cup B)\cap(C\cup
D)$.
\vs{-2}
\begin{itemize}
\item[1.] Inclusion: $(A,B)\subseteq
(C,D)$ iff $A\subseteq C$ and $A\cup B= C\cup D$.
\vs{-2}
\item[2.] Intersection: $(A,B)\cap(C,D)\stackrel{def}{=} 
(A\cap C,(B\cup D)\cap E)$.
\vs{-2}
\item[3.] Union: $(A,B)\cup(C,D) \stackrel{def}{=} 
((A\cup C)\cap E,B\cap D)$.
\vs{-2}
\item[4.] Complementation: $\overline{(A,B)} 
\stackrel{def}{=} (B,A)$.
\end{itemize}
\vs{-2}

Now, we focus on classes of partial decision problems. To denote such
a class, we use the special notation $\quantset{\CC}$, whose asterisk
signifies the deviation from total decision problems. The {\em partial
classical part} of a partial decision problem $(A,B)$ is
$(\check{A},\check{B})$. When $\check{A}\cup\check{B}=\Sigma^{*}$, we
call $(\check{A},\check{B})$ the {\em total classical part} of $(A,B)$
and simply write $\check{A}$ instead of $(\check{A},\check{B})$ as
before. Notationally, let $\CC$ denote the collection of {\em total}
classical parts in $\quantset{\CC}$. We call $\CC$ the {\em total
classical part} of $\quantset{\CC}$.

For later use, we expand $\bqp$ to the class of partial decision
problems. In a similar fashion, we can expand other classes, such as
$\nqp$ and $\pqp$ \cite{Yam99b}.

\begin{definition}
Let $a,b$ be any two functions from $\nat$ to $[0,1]$ such that
$a(n)+b(n)=1$ for all $n\in\nat$.  A partial decision problem $(A,B)$
is in $\quantset{\bqp}(a,b)$ if there exists a quantum function
$f\in\quantset{\sharpqp}$ such that, for every
$\qubit{\vec{\phi}}\in\qustrings_{\infty}^{*}$, (i) if
$\qubit{\vec{\phi}}\in A$ then $f(\qubit{\vec{\phi}})\geq
a(\ell(\qubit{\vec{\phi}}))$ and (ii) if $\qubit{\vec{\phi}}\in B$
then $f(\qubit{\vec{\phi}})\leq b(\ell(\qubit{\vec{\phi}}))$. For
simplicity, write $\quantset{\bqp}$ for $\quantset{\bqp}(3/4,1/4)$.
\end{definition}

It is important to note that the total classical part of
$\quantset{\bqp}$ coincides with the standard definition of $\bqp$,
e.g., given in \cite{BV97}.  Since the duplication of a quantum input
is available for free of charge, we can perform a standard
majority-vote algorithm for a set in $\quantset{\bqp}$ to amplify its
success probability. Therefore, we obtain
$\quantset{\bqp}=\quantset{\bqp}(1-2^{-p(n)},2^{-p(n)})$ for any
polynomial $p$.

\section{The $\exists^{\mathrm{Q}}$-Operator 
and the $\forall^{\mathrm{Q}}$-Operator}\label{sec:quantifier}

\n The process of defining a new complexity class $\DD$ from a 
basis class $\CC$ can be naturally viewed as an application of an
operator, which maps $\CC$ to $\DD$. As seen in Section
\ref{sec:quantum-NP}, the $\exists$-operator over classical sets 
is an abstraction of nondeterministic computation (as in
$\np=\exists\cdot\p$) and its complement is called the {\em
$\forall$-operator}. First, we generalize these operators to the ones
whose scopes are classes of partial decision problems.
 
\begin{definition}\label{def:exists-operator}
Let $\quantset{\CC}$ be any quantum complexity class of partial
decision problems. A partial decision problem $(A,B)$ is in
$\quantset{\exists}\cdot\quantset{\CC}$ if there exist a polynomial
$p$ and a partial decision problem $(C,D)$ in $\quantset{\CC}$ such
that, for all vectors $\qubit{\vec{\phi}}\in\qustrings_{\infty}^{*}$,
\vs{-2}
\begin{itemize}
\item[i)] if $\qubit{\vec{\phi}}\in A$ then 
$\exists x\in\Sigma^{p(\ell(\qubit{\vec{\phi}}))}
[(\qubit{x},\qubit{\vec{\phi}})\in C]$ and
\vs{-2}
\item[ii)] if $\qubit{\vec{\phi}}\in B$ then 
$\forall x\in \Sigma^{p(\ell(\qubit{\vec{\phi}}))}
[(\qubit{x},\qubit{\vec{\phi}})\in D]$.
\end{itemize}
\vs{-2}
The class $\quantset{\forall}\cdot\quantset{\CC}$ is defined similarly
by exchanging the roles of the quantifiers in conditions i) and ii). In
accordance to the standard notation, $\exists\cdot\quantset{\CC}$ and
$\forall\cdot\quantset{\CC}$ denote the total classical parts of
$\quantset{\exists}\cdot\quantset{\CC}$ and
$\quantset{\forall}\cdot\quantset{\CC}$, respectively.
\end{definition}

The class $\qma$ uses a quantum quantifier, whose scope is qustrings
of polynomial size instead of classical strings of polynomial length.
Generalizing such a quantum quantifier, 
we introduce a quantum analogue of
the $\exists$- and $\forall$-operators as follows. Our approach is
quite different from that of Green \etalc~\cite{GHMP02}, who defined
the N-operator as an abstraction of $\nqp$.

\begin{definition}\label{def:QE-operator}
Let $\quantset{\CC}$ be a quantum complexity class of partial decision
problems. A partial decision problem $(A,B)$ is in
$\quantset{\qe}\quantset{\CC}$ if there exist a polynomial $p$ and a
partial decision problem $(C,D)\in\quantset{\CC}$ such that, for every
$\qubit{\vec{\phi}}\in\qustrings_{\infty}^{*}$,
\vs{-2}
\begin{itemize}
\item[i)] if $\qubit{\vec{\phi}}\in A$ then 
$\exists \qubit{\psi}\in\qustrings_{p(\ell(\qubit{\vec{\phi}}))} 
[(\qubit{\psi},\qubit{\vec{\phi}})\in C]$ and
\vs{-2}
\item[ii)] if $\qubit{\vec{\phi}}\in B$ then 
$\forall \qubit{\psi}\in\qustrings_{p(\ell(\qubit{\vec{\phi}}))} 
[(\qubit{\psi},\qubit{\vec{\phi}})\in D]$.
\end{itemize}
\vs{-2}
Similarly, the class $\quantset{\qa}\quantset{\CC}$ is defined by
exchanging the roles of quantifiers in conditions i) and ii)
above. The notations $\qe\quantset{\CC}$ and $\qa\quantset{\CC}$
denote the total classical parts of $\quantset{\qe}\quantset{\CC}$ and
$\quantset{\qa}\quantset{\CC}$, respectively. More generally, write
$\quantset{\exists^{\mathrm{Q}}_{1}}\cdot\quantset{\CC}$ for
$\quantset{\qe}\quantset{\CC}$ and recursively define
$\quantset{\exists^{\mathrm{Q}}_{m+1}}\cdot\quantset{\CC}$ as
$\quantset{\qe}(\quantset{\exists^{\mathrm{Q}}_{m}}
\cdot\quantset{\CC})$. Similarly,
$\quantset{\forall^{\mathrm{Q}}_{m+1}}\cdot\quantset{\CC}$ is defined.
\end{definition}

Obviously, if $\quantset{\CC}\subseteq\quantset{\DD}$ then
$\qe\quantset{\CC}\subseteq\qe\quantset{\DD}$ and
$\quantset{\qe\quantset{\CC}}\subseteq\quantset{\qe\quantset{\DD}}$.

We next show that the $\exists^{\mathrm{Q}}$- and
$\forall^{\mathrm{Q}}$-operators indeed expand the classical
$\exists$- and $\forall$-operators, respectively. Proving this claim,
however, requires underlying class $\quantset{\CC}$ to satisfy a
certain condition, which is given in the following definition.

\begin{definition}
1. A quantum set $B\subseteq\qustrings_{\infty}^{*}$ is called {\em
classically separable} if the following condition holds: for every
$m,n\in\nat^+$ and every $\qubit{\vec{\phi}}\in\qustrings_{n}^{m}$, if
either $\measure{\vec{x}}{\vec{\phi}}=0$ or
$(\qubit{\vec{x}},\qubit{\vec{\psi}})\in B$ for all
$\vec{x}\in(\Sigma^{n})^{m}$, then
$(\qubit{\vec{\phi}},\qubit{\vec{\psi}})\in B$.

2. A quantum complexity class $\quantset{\CC}$ of partial decision
problems is said to be {\em classically simulatable} if, for every
partial decision problem $(A,B)\in\quantset{\CC}$, there exist a
partial decision problem $(C,D)\in\quantset{\CC}$ such that (i) $C$
and $D$ are classically separable and (ii) for all $m,n\in\nat^{+}$
and all $\vec{x}\in(\Sigma^n)^{m}$,
$(\qubit{\vec{x}},\qubit{\vec{\psi}})\in A$ $\IFF$
$(\qubit{\vec{x}},\qubit{\vec{\psi}})\in C$ and
$(\qubit{\vec{x}},\qubit{\vec{\psi}})\in B$ $\IFF$
$(\qubit{\vec{x}},\qubit{\vec{\psi}})\in D$.
\end{definition}

The above notion stems from the proof of $\qma$ containing $\np$. The
classes of partial decision problems dealt with in this paper are
indeed classically simulatable.

\begin{lemma}\label{lemma:simulatable}
If a quantum complexity class $\quantset{\CC}$ of partial decision
problems is classically simulatable, then
$\quantset{\exists}\cdot\quantset{\CC}
\subseteq\quantset{\qe}\quantset{\CC}$ and
$\quantset{\forall}\cdot\quantset{\CC}
\subseteq\quantset{\qa}\quantset{\CC}$.
\end{lemma}

The proof of the first claim of Lemma \ref{lemma:simulatable} easily
follows from the definition of classical-simulatability. The second
claim comes from the fact that if $\quantset{\CC}$ is classically
simulatable then so is $\co\quantset{\CC}$, where $\co\quantset{\CC}$
denotes the collection of partial decision problems whose complements
belong to $\quantset{\CC}$.

\section{The Quantum Polynomial Hierarchy}

\n The classical $\exists$- and $\forall$-operators are 
useful tools to expand complexity classes. In the early 1970s, 
Meyer and Stockmeyer
\cite{MS72} introduced the {\em polynomial hierarchy}, whose
components are obtained from $\p$ with alternating applications of
these operators; namely, $\sigmah{k+1}=\exists\cdot\pih{k}$ and
$\pih{k+1}=\forall\cdot\sigmah{k}$ for each level $k\in\nat^{+}$.  The
polynomial hierarchy has continued to be a center of research in
complexity theory. The introduction of the $\exists^{\mathrm{Q}}$- and
$\forall^{\mathrm{Q}}$-operators enables us to consider a quantum
analogue of the polynomial hierarchy and explore its structural
properties in light of the strength of quantum computability. We call
this new hierarchy the {\em quantum polynomial hierarchy}
(QP-hierarchy, for short).  The basis of the QP hierarchy is
$\quantset{\bqp}$ opposed to $\p$ since two-sided bounded-error
computations are more realistic in the quantum setting.  Each level of
the QP hierarchy is obtained from its lower level by a {\em finite
number} of applications of the same quantum operator (either
$\exists^{\mathrm{Q}}$- or $\forall^{\mathrm{Q}}$-operators). Although
any repetition of the same classical operator has no significance in
the polynomial hierarchy, as Kobayashi \etalc~\cite{KMY01} pointed
out, there might be a potentially essential difference between a
single quantum quantifier and multiple quantum quantifiers of the same
type.  The precise definition of the QP hierarchy is given as follows.

\begin{definition}\label{def:QPH}
Let $a,b$ be two functions from $\nat$ to $[0,1]$ such that
$a(n)+b(n)=1$ for all $n\in\nat$. Let $k\in\nat$ and
$m\in\nat^{+}$. The {\em quantum polynomial hierarchy} (QP hierarchy,
for short) constitutes the following complexity classes of partial
decision problems.
\vs{-2}
\begin{enumerate}
\item[i)] $\quantset{\sigmaqh{0,m}}(a,b)= \quantset{\piqh{0,m}}(a,b)= 
\quantset{\bqp}(a,b)$.
\vs{-2}
\item[ii)] $\quantset{\sigmaqh{k+1,m}}(a,b)= 
\quantset{\exists^{\mathrm{Q}}_{m}}\cdot\quantset{\piqh{k,m}}(a,b)$.
\vs{-2}
\item[iii)] $\quantset{\piqh{k+1,m}}(a,b)=
\quantset{\forall^{\mathrm{Q}}_{m}}\cdot \quantset{\sigmaqh{k,m}}(a,b)$.
\end{enumerate}
\vs{-2}
Let $\quantset{\sigmaqh{k}}(a,b) = \bigcup_{m\geq
1}\quantset{\sigmaqh{k,m}}(a,b)$ and $\quantset{\piqh{k}}(a,b) =
\bigcup_{m\geq 1}\quantset{\piqh{k,m}}(a,b)$. Furthermore, let
$\quantset{\qph}_{m}(a,b) =
\bigcup_{k\geq0}(\quantset{\sigmaqh{k,m}}(a,b)
\cup\quantset{\piqh{k,m}}(a,b))$
and $\quantset{\qph}(a,b)= \bigcup_{m\geq1}
\quantset{\qph_m}(a,b)$.
Their total classical parts are denoted
(without asterisks) $\sigmaqh{k}(a,b)$, $\piqh{k}(a,b)$, and
$\qph(a,b)$.
\end{definition}

For brevity, we write $\quantset{\sigmaqh{k}}$ for
$\quantset{\sigmaqh{k}}(3/4,1/4)$, $\quantset{\piqh{k}}$ for
$\quantset{\piqh{k}}(3/4,1/4)$, and $\quantset{\qph}$ for
$\quantset{\qph}(3/4,1/4)$. Likewise, we can define their total
classical parts $\sigmaqh{k}$, $\piqh{k}$, and $\qph$. The choice of
the value $(3/4,1/3)$ is artificial; however, a standard majority-vote
algorithm can amplify $(3/4,1/4)$ to $(1-2^{-p(n)},2^{-p(n)})$ for an
arbitrary polynomial $p$. Due to our general framework, it is likely
that $\sigmaqh{1}$ is strictly larger than
$\bigcup_{k\geq1}\qma(k)$. {}From this reason, $\sigmaqh{1}$ can be
regarded as Quantum NP, as discussed in Section \ref{sec:quantum-NP}.

Several alternative definitions of the QP hierarchy are
possible. Here, we present three alternatives.  The first one uses the
function class $\quantset{\qoptsharpsigmaqh{k,m}}$---a generalization
of $\qoptsharpqp$ in \cite{Yam02}---introduced as follows: a quantum
function $f$ from $\qustrings_{\infty}^{*}$ to $[0,1]$ is in
$\quantset{\qoptsharpsigmaqh{k,m}}$ if there exist a polynomial $p$
and a quantum function $g\in \quantset{\sharpqp}$ such that, for every
$\qubit{\vec{\phi}}\in\qustrings_{\infty}^{*}$,
\[
f(\qubit{\vec{\phi}})=
\sup_{\qubit{\vec{\psi}_1}}\inf_{\qubit{\vec{\psi}_2}}\cdots
\mathrm{opr}^{(k)}_{\qubit{\vec{\psi}_k}} 
\{g(\qubit{\vec{\phi}},\qubit{\vec{\psi}_1},
\qubit{\vec{\psi}_2},\ldots,\qubit{\vec{\psi}_k})\},
\]
where $\mathrm{opr}^{(k)}=\sup$ if
$k$ is odd and $\mathrm{opr}^{(k)}=\inf$ otherwise, and each
$\qubit{\vec{\psi}_i}$ is an $m$-tuple 
$(\qubit{\psi_{i,1}},\qubit{\psi_{i,2}},
\ldots,\qubit{\psi_{i,m}})$ with each $\qubit{\psi_{i,j}}$ 
running over all qustrings of size $p(\ell(\qubit{\vec{\phi}}))$. The
class $\quantset{\qoptsharpsigmaqh{k,m}}$ gives a succinct way to define
the $k$th level of the QP hierarchy.

\begin{lemma}\label{lemma:qoptsharpqp}
Let $k,m\geq1$ and let $a,b$ be any two functions from $\nat$ to
$[0,1]$ satisfying $a(n)+b(n)=1$ for all $n\in\nat$. A partial
decision problem $(A,B)$ is in $\quantset{\sigmaqh{k,m}}(a,b)$ iff
there exists a quantum function $f$ in
$\quantset{\qoptsharpsigmaqh{k,m}}$ such that, for every
$\qubit{\vec{\phi}}\in\qustrings_{\infty}^{*}$, (i) if
$\qubit{\vec{\phi}}\in A$ then $f(\qubit{\vec{\phi}})\geq
a(\ell(\qubit{\vec{\phi}}))$ and (ii) if $\qubit{\vec{\phi}}\in B$
then $f(\qubit{\vec{\phi}})\leq b(\ell(\qubit{\vec{\phi}}))$.
\end{lemma}

In the second alternative definition, we use vectors whose components
are described by classical strings. For each $n,r\in\nat^{+}$, denote
by $\tilde{\qustrings}_n(r)$ the collection of all vectors
$\qubit{\phi}$ (not necessarily elements in a Hilbert space) such that
$\qubit{\phi}$ has the form $\sum_{s:|s|=n}\alpha_s\qubit{s}$, where
each complex number $\alpha_s$ is expressed as a pair of two binary
fractions of $r$ bits. If $r\geq\log(1/\epsilon)+n+2$ for
$\epsilon>0$, such $\qubit{\phi}$ satisfies
$|\sum_{s:|s|=n}|\alpha_s|^2-1|\leq
\epsilon$. Note that any element in $\tilde{\qustrings}_{n}$ can be
expressed as a binary string of length $r2^{n+1}$ (since each
$\alpha_s$ needs $2r$ bits and we have exactly $2^n$ such
$\alpha_s$'s). Thus, the cardinality of $\tilde{\qustrings}_{n}$ is
$2^{r2^{n+1}}$. For our purpose, we allow each
$\quantset{\sharpqp}$-function to take any vector in
$\tilde{\qustrings}_{n}(r)$ as its input.

\begin{lemma}\label{lemma:input-approximation}
Let $k,m\in\nat^{+}$. A partial decision problem $(A,B)$ is in
$\quantset{\sigmaqh{k,m}}$ iff there exist a polynomial $p$ and a
quantum function $f\in\quantset{\sharpqp}$ such that, for every series
of qustrings $\qubit{\vec{\phi}}$ in $\qustrings_{\infty}^{*}$,
\vs{-2}
\begin{itemize}
\item[i)] if $\qubit{\vec{\phi}}\in
A$ then
$\exists\qubit{\vec{\xi_1}}\forall\qubit{\vec{\xi_2}}\cdots
Q_k\qubit{\vec{\xi_k}}[ f(\qubit{\vec{\phi}},
\qubit{\vec{\xi_1}},\qubit{\vec{\xi_2}},\ldots,\qubit{\vec{\xi_k}}) 
\geq 3/4]$ and
\vs{-2}
\item[ii)] if $\qubit{\vec{\phi}}\in B$ then
$\forall\qubit{\vec{\xi_1}}\exists\qubit{\vec{\xi_2}}
\cdots \overline{Q}_k\qubit{\vec{\xi_k}}[
 f(\qubit{\vec{\phi}},\qubit{\vec{\xi_1}},
\qubit{\vec{\xi_2}},\ldots,\qubit{\vec{\xi_k}})\leq 1/4]$,
\end{itemize}
\vs{-2}
where each variable $\qubit{\vec{\xi_i}}$ runs over all series of $m$
vectors in $\tilde{\qustrings}_{p(\ell(\qubit{\vec{\phi}}))}
(3p(\ell(\qubit{\vec{\phi}})))$, $Q_k=\forall$ if $k$ is even and
$Q_k=\exists$ otherwise, and $\overline{Q}_k$ is the opposite
quantifier of $Q_k$.
\end{lemma}

The last alternative definition is much more involved and we need
extra notions and notation. Firstly, we give a method of translating a
qustring $\qubit{\phi}$ into a series of unitary matrices that
generate $\qubit{\phi}$. Let $\complex$ be the set of all complex
numbers, $I$ the $2\times2$ identity matrix, and $\lambda$ the empty
string.  Fix $n\in\nat$ and $\qubit{\phi}\in\qustrings_{n+1}$ and
assume that $\qubit{\phi}=\sum_{s:|s|=n+1}\gamma_{s}\qubit{s}$, where
each $\gamma_s$ is in $\complex$. For each $s\in\Sigma^{\leq n}$, set
$g_s=
\sqrt{\sum_{t}|\gamma_{s0t}|^2 + \sum_{t}|\gamma_{s1t}|^2}$ and 
define a $2\times2$ matrix $U^{(s)}$ as follows: let
$U^{(s)}\qubit{b} = (\sqrt{\sum_{t}|\gamma_{s0t}|^2}/g_{s})\qubit{0} +
(-1)^b(\sqrt{\sum_{t}|\gamma_{s1t}|^2}/g_{s})\qubit{1}$ if $|s|<n$;
otherwise, let $U^{(s)}\qubit{b} = (\gamma_{s0}/g_{s})\qubit{0} +
(-1)^b(\gamma_{s1}/g_{s})\qubit{1}$.  The series
$\UU=\pair{U^{(s)}\mid s\in\Sigma^{\leq n}}$ is called the {\em
generator} of $\qubit{\phi}$ since $\qubit{\phi}=U_{n}U_{n-1}\cdots
U_0\qubit{0^{n+1}}$, where $U_0= U^{(\lambda)}\otimes I^{\otimes{n}}$
and $U_k=\sum_{s:|s|=k}\ket{s}\bra{s}\otimes U^{(s)}\otimes
I^{\otimes{n-k}}$ for each $k$, $1\leq k\leq n$.

Secondly, we consider a good approximation of a given generator.  For
any $2\times2$ matrix $U=(u_{ij})_{1\leq i,j\leq 2}$ on $\complex$ and
any $\epsilon>0$, $\tilde{U}=(\tilde{u}_{ij})_{1\leq i,j\leq 2}$ is
called the {\em $\epsilon$-fragment} of $U$ if each $\tilde{u}_{ij}$
represents the first $\ceilings{\log(1/\epsilon)}$ bits of the
infinite binary fractions of the real and imaginary parts of $u_{ij}$
(so that $|u_{ij}-\tilde{u}_{ij}|\leq 2\epsilon$). In this case,
$\tilde{U}$ satisfies $\|U-\tilde{U}\|\leq 4\epsilon$.  If
$\tilde{U}^{(s)}$ is the $\epsilon$-fragment of $U^{(s)}$ for all
$s\in\Sigma^{\leq n}$, the series $\tilde{\UU} \stackrel{def}{=}
\pair{\tilde{U}^{(s)}\mid s\in\Sigma^{\leq n}}$ is also called the 
{\em $\epsilon$-fragment} of $\UU$.  We assume a natural encoding
scheme of $\tilde{\UU}$ into oracle $\pair{\tilde{\UU}}$ so that
$\tilde{\UU}$ can be retrieved by $O(2^n\log(1/\epsilon))$ queries to
oracle $\pair{\tilde{\UU}}$.

\begin{lemma}\label{lemma:oracle}
There exists a well-formed QTM $M_0$ that satisfies the following
condition: for every $\epsilon>0$, every $n\in\nat$, and every
generator $\UU$ of a qustring $\qubit{\phi}\in\qustrings_{n+1}$, if
$\tilde{\UU}$ is the $\epsilon 2^{-n-4}$-fragment of $\UU$, then $M_0$
with oracle $\pair{\tilde{\UU}}$ halts on input $\qubit{0^{n+1}}$ in
time polynomial in $1/\epsilon$ and $n$ and satisfies
$\|\ket{\phi}\bra{\phi}-\rho\|_{\mathrm{tr}}\leq
\epsilon$, where $\rho$ is the density matrix obtained from the final
superposition of $M_0$ by tracing out all but the output-tape content
and $\|A\|_{\mathrm{tr}}$ denotes the trace of $\sqrt{A^{\dagger}A}$.
\end{lemma}

\begin{proofsketch}
The desired $M_0$ works as follows: at step 0, write $\qubit{0^{n+1}}$
in the work tape. Let $s0^{n-k+1}$ be the string written in the work
tape after step $k-1$. At step $k$, make appropriate queries to oracle
$\pair{\tilde{\UU}}$ to realize quantum gate $G^{(s)}$ that simulates
$\tilde{U}^{(s)}$ with accuracy at most $\delta$ (i.e.,
$\|G^{(s)}-\tilde{U}^{(s)}\|\leq\delta$). Then, apply
$\ket{s}\bra{s}\otimes G^{(s)}\otimes I^{\otimes{n-k}}$ (or
$G^{(\lambda)}\otimes I^{\otimes{n}}$ if $s=\lambda$) to
$\qubit{s0^{n-k+1}}$.
\end{proofsketch}

Finally, the third alternative definition of the QP hierarchy is
given in Lemma \ref{lemma:oracle-charact}. A merit of Lemma
\ref{lemma:oracle-charact} is no need of the duplication of quantum
inputs given to underlying QTMs.  Note that Lemma
\ref{lemma:oracle-charact} can be further generalized to non-generators.

\begin{lemma}\label{lemma:oracle-charact}
Let $k\geq1$. For any classical set $A\subseteq\Sigma^*$, $A$ is in
$\sigmaqh{k,1}$ iff there exist two polynomials $p,q$ and a
polynomial-time well-formed oracle QTM $M$ such that, for all
$x\in\Sigma^*$,
\vs{-2}
\begin{itemize}
\item[i)] if $x\in A$ then $\exists \UU_1\forall \UU_2
\cdots Q_k\UU_k[\prob_{M}[M^{\pair{\tilde{\UU}_1,\tilde{\UU}_2,
\ldots,\tilde{\UU}_k}}(x)=1]\geq 3/4]$ and
\vs{-2}
\item[ii)] if $x\not\in A$ then $\forall \UU_1\exists
 \UU_2\cdots \overline{Q}_k\UU_k[\prob_{M}[M^{\pair{\tilde{\UU}_1,
\tilde{\UU}_2,\ldots,\tilde{\UU}_k}}(x)=1]\leq 1/4]$,
\end{itemize}
\vs{-2}
where $Q_k=\forall$ if $k$ is even and $Q_k=\exists$ otherwise,
$\overline{Q}_k$ is the opposite quantifier of $Q_k$, each variable
$\UU_i$ runs over all generators of qustrings of size $p(|x|)$, each
$\tilde{\UU}_i$ is the $2^{-q(|x|)}$-fragment of $\UU_i$, and $M$ on
input $x$ behaves as follows: whenever it makes a query, it writes
$\qubit{1^i}$ in an query tape and runs $M_0$ (defined in Lemma
\ref{lemma:oracle}) on input $\qubit{0^{p(|x|)}}$ with oracle
$\pair{\tilde{\UU}_i}$.
\end{lemma}

\section{Fundamental Properties of the QP 
Hierarchy}\label{sec:properties}

\n Although the QP hierarchy looks more complex than its
classical counterpart, the QP hierarchy shares many fundamental
properties with the polynomial hierarchy. In the next proposition,
we present without proofs a short list of fundamental
properties of the QP hierarchy.

\begin{proposition}\label{prop:properties}
\begin{itemize}
\item[1.] For every $k\in\nat$, 
$\co\quantset{\sigmaqh{k}}=\quantset{\piqh{k}}$
and $\co\quantset{\piqh{k}}=\quantset{\sigmaqh{k}}$.
\vs{-2}
\item[2.] For each $k\in\nat$, $\quantset{\sigmaqh{k}}$ 
and $\quantset{\piqh{k}}$
are closed under intersection and union.
\vs{-2}
\item[3.] For each $k\in\nat$,
$\quantset{\sigmaqh{k}}\cup\quantset{\piqh{k}}
\subseteq\quantset{\sigmaqh{k+1}}\cap\quantset{\piqh{k+1}}$.
\vs{-2}
\item[4.] For every $k\in\nat^{+}$,
$\bigcup_{m>0}\quantset{\exists}^{\mathrm{Q}}_{m}
\cdot(\quantset{\sigmaqh{k}}\cap\quantset{\piqh{k}})
= \quantset{\sigmaqh{k}}$.
\vs{-2}
\item[5.] Let $k\in\nat^{+}$. If 
$\quantset{\Sigma}^{\mathrm{QP}}_{k}=
\quantset{\Pi}^{\mathrm{QP}}_{k}$ then
$\quantset{\Sigma}^{\mathrm{QP}}_{k}= \quantset{\qph}$.
\vs{-2}
\item[6.] For each $k\in\nat^{+}$, 
$\quantset{\exists}\cdot\quantset{\sigmaqh{k}}=
\quantset{\qe}\quantset{\Sigma}^{\mathrm{QP}}_{k} =
\quantset{\sigmaqh{k}}$. 
\end{itemize}
\end{proposition}

Of the above items, item 6 is specifically meant for the QP hierarchy
and requires the classical-simulatability of the QP hierarchy, which
is shown below.

\begin{lemma}\label{lemma:sigma-pi-simulatable}
For each $k\in\nat$, $\quantset{\sigmaqh{k}}$ and
$\quantset{\piqh{k}}$ are classically simulatable.
\end{lemma}

\begin{proof}
Since the base case $k=0$ is easy, we skip this case and prove the
general case $k>0$. Let $m\in\nat^{+}$ and $(A,B)$ be any partial
decision problem in $\quantset{\sigmaqh{k,m}}$.  There exists a
function $f\in\quantset{\qoptsharpsigmaqh{k,m}}$ that satisfies Lemma
\ref{lemma:qoptsharpqp} for $(A,B)$.  Assume that $f$ has the form
$f(\qubit{\vec{\phi}},\qubit{\vec{\psi}})=
\sup_{\qubit{\vec{\xi}_1}}\inf_{\qubit{\vec{\xi}_2}}\cdots
\mathrm{opr}^{(k)}_{\qubit{\vec{\xi}_k}} 
\{h(\qubit{\vec{\xi}_1},\qubit{\vec{\xi}_2},
\ldots,\qubit{\vec{\xi}_k},\qubit{\vec{\phi}},\qubit{\vec{\psi}})\}$
for a certain quantum function $h$ in $\quantset{\sharpqp}$. For
brevity, write $\qubit{\Xi}$ for
$(\qubit{\vec{\xi}_1},\ldots,\qubit{\vec{\xi}_k})$. Letting
$h'(\qubit{\Xi},\qubit{\vec{\phi}},\qubit{\vec{\psi}}) =
\sum_{\vec{x}}|\measure{\vec{x}}{\vec{\phi}}|^2h(\qubit{\Xi},
\qubit{\vec{x}},\qubit{\vec{\psi}})$, we define $g$ as
$g(\qubit{\vec{\phi}},\qubit{\vec{\psi}})=
\sup_{\qubit{\vec{\xi}_1}}\inf_{\qubit{\vec{\xi}_2}}\cdots
\mathrm{opr}^{(k)}_{\qubit{\vec{\xi}_k}} 
\{h'(\qubit{\vec{\xi}_1},\qubit{\vec{\xi}_2},\ldots,
\qubit{\vec{\xi}_k},\qubit{\vec{\phi}},\qubit{\vec{\psi}})\}$. 
Note that $g(\qubit{\vec{x}},\qubit{\vec{\psi}})\leq
g(\qubit{\vec{\phi}},\qubit{\vec{\psi}})$ if
$\measure{\vec{x}}{\vec{\phi}}\neq0$, since
$h'$ satisfies 
$h'(\qubit{\Xi},\qubit{\vec{x}},\qubit{\vec{\psi}})\leq
h'(\qubit{\Xi},\qubit{\vec{\phi}},\qubit{\vec{\psi}})$ for any
$\qubit{\Xi}$.  It follows that, for every $\vec{x}$,
$g(\qubit{\vec{x}},\qubit{\vec{\psi}}) =
f(\qubit{\vec{x}},\qubit{\vec{\psi}})$.

To complete the proof, defining
$C=\{(\qubit{\vec{\phi}},\qubit{\vec{\psi}})\mid
g(\qubit{\vec{\phi}},\qubit{\vec{\psi}})\geq 3/4\}$ and
$D=\{(\qubit{\vec{\phi}},\qubit{\vec{\psi}})\mid
g(\qubit{\vec{\phi}},\qubit{\vec{\psi}})\leq 1/4\}$, we show that $C$
and $D$ are classically separable. Let $\qubit{\vec{\phi}}$ be fixed
arbitrarily. Assume that
$\forall\vec{x}[\measure{\vec{x}}{\vec{\phi}}=0 \vee
(\qubit{\vec{x}},\qubit{\vec{\psi}})\in C]$. We want to show that
$(\qubit{\vec{\phi}},\qubit{\vec{\psi}})\in C$. Assume otherwise. We
then have $g(\qubit{\vec{\phi}},\qubit{\vec{\psi}})<3/4$. Take any
element $\vec{x}$ such that $\measure{\vec{x}}{\vec{\phi}}\neq0$. It
follows that $g(\qubit{\vec{x}},\qubit{\vec{\psi}})\leq
g(\qubit{\vec{\phi}},\qubit{\vec{\psi}})<3/4$, which implies
$(\qubit{\vec{x}},\qubit{\vec{\psi}})\not\in C$, a contradiction.  The
case for $D$ is similar. Therefore, $\quantset{\sigmaqh{k}}$ is
classically simulatable.
\end{proof}

Next, we give rudimentary but meritorious upper and lower bounds of
the QP hierarchy.  The {\em exponential hierarchy} consists of the
following classes: $\deltaexph{0}=\sigmaexph{0}=\piexph{0}=\dexp$
($=\dtime{2^{n^{O(1)}}}$), $\deltaexph{k}=\dexp^{\sigmah{k}}$,
$\sigmaexph{k}=\nexp^{\sigmah{k-1}}$, and
$\piexph{k}=\co\sigmaexph{k}$ for every $k\in\nat^{+}$. Let $\exph$
denote the union of $\sigmaexph{k}$ for all $k\in\nat$.  We show that
$\ph\subseteq\qph\subseteq\exph$.

\begin{theorem}\label{theorem:upper-bound}
For each $k>0$, $\sigmah{k}\subseteq \sigmaqh{k}
\subseteq\sigmaexph{k}$. Thus, $\ph\subseteq \qph\subseteq\exph$.
\end{theorem}

Theorem \ref{theorem:upper-bound} yields the following collapse: 
if $\pih{k}\subseteq\sigmaqh{k}$ then $\ph\subseteq\sigmaqh{k}$.

\begin{proofof}{Theorem \ref{theorem:upper-bound}}
We show the first inclusion that $\sigmah{k}\subseteq \sigmaqh{k}$.
The proof is done by induction on $k\geq0$. The base case $k=0$
follows from $\p\subseteq\bqp$. Let $k>0$. By the induction
hypothesis, it follows that $\sigmah{k-1}\subseteq\sigmaqh{k-1}$,
which further implies $\pih{k-1}\subseteq\piqh{k-1}$. Since
$\quantset{\piqh{k-1}}$ is classically simulatable (Lemma
\ref{lemma:sigma-pi-simulatable}), Lemma \ref{lemma:simulatable}
yields $\exists\cdot\quantset{\piqh{k-1}}
\subseteq\qe\quantset{\piqh{k-1}}$. Thus, 
\[
 \sigmah{k}=\exists\cdot\pih{k-1}
 \subseteq\exists\cdot\quantset{\piqh{k-1}}
\subseteq \bigcup_{m>0}
\exists^{\mathrm{Q}}_{m}\cdot\quantset{\piqh{k-1}} =\sigmaqh{k}.
\]

The second inclusion $\sigmaqh{k}
\subseteq\sigmaexph{k}$ follows from Lemma
\ref{lemma:input-approximation}. Let $m\in\nat^{+}$ and let 
$A$ be any set in $\sigmaqh{k,m}$. Take a polynomial $p$ and a quantum
function $f\in\quantset{\sharpqp}$ guaranteed by Lemma
\ref{lemma:input-approximation} for $(A,\overline{A})$.  We construct
an alternating TM $N$ as follows: on input $x$, start with an
$\exists$-state, generate $k$ vectors
$\qubit{\vec{\xi_1}},\qubit{\vec{\xi_2}},\ldots,\qubit{\vec{\xi_k}}$
in $(\tilde{\qustrings}_{p(|x|)}(3p(|x|)))^m$ by alternately entering
$\forall$- and $\exists$-states, and check if
$f(\qubit{x},\qubit{\vec{\xi_1}},\qubit{\vec{\xi_2}},
\ldots,\qubit{\vec{\xi_k}})\geq 3/4$. This last check is
done in exponential time since $f$ runs in time polynomial in $|x|$.
Since any exponential-time alternating TM with $k$-alternation
starting with the $\exists$-state is known to characterize
$\sigmaexph{k}$, $A$ belongs to $\sigmaexph{k}$.
\end{proofof}

In the end of this section, we discuss the issue of complete
problems. Each level of the polynomial hierarchy is known to have
complete problems. Unfortunately, it is believed that classes like
$\bqp$ and $\qma$ lack such complete problems because of their
acceptance criteria. Dealing with partial decision problems, however,
allows us to go around this difficulty. (See,
\eg \cite{DK00} for the NP-completeness for classical partial decision
problems.) With an appropriate modification of classical completeness
proofs, we can show that each level of the QP hierarchy indeed has a
``complete'' partial decision problem (under a deterministic
reduction). An important open problem is to find natural complete
partial decision problems for each level of the QP hierarchy.

\section{Relativized QP Hierarchies}\label{sec:oracle}

\n We have introduced a quantum analogue of the 
polynomial hierarchy and explored its basic properties and its
relationship to the polynomial hierarchy.  In this last section, we
give simple relativized results related to the QP hierarchy. The {\em
relativized QP hierarchy relative to} oracle $A$,
$\{\sigmaqh{k}(A),\piqh{k}(A)\mid k\in\nat\}$, is obtained simply by
changing the basis class $\quantset{\sharpqp}$ to its relativized
version $\quantset{\sharpqp}^A$.

\begin{proposition}\label{prop:oracle-separation}
1. There exists a recursive oracle $A$ such that $\p^A=\ph^A=\qph^A$.

2. There exists a recursive oracle $B$ such that
$\sigmaqh{0}(B)\neq\sigmaqh{1}(B)\neq\sigmaqh{2}(B)$.

3. There exists a recursive oracle $C$ such that
$\sigmah{k}(C)\neq\sigmaqh{k}(C)$ for all $k\in\nat^{+}$.
\end{proposition}

For the first claim of Proposition \ref{prop:oracle-separation}, it
suffices to construct $A$ such that $\p^A=\sigmaqh{1}(A)$ since this
yields $\sigmaqh{1}(A)=\piqh{1}(A)$, which further implies
$\sigmaqh{1}(A)=\qph^A$.  The desired set $A$ is built by stages: at
each stage, pick one relativized $\sharpqp$-function $f$ and encode
its outcome into one string that cannot be queried by $f$. This is
possible because an oracle QTM that witnesses $f$ runs in polynomial
time.

The second claim of Proposition \ref{prop:oracle-separation} follows
from a generalization of the result
$\co\up^B\nsubseteq\bigcup_{k\geq1}\qma(k)^B$ \cite{KMY01}. In fact,
we can show by modifying Ko's argument
\cite{Ko90} a slightly stronger result: there exists an oracle $B$
satisfying that $\co\up^B\nsubseteq\sigmaqh{1}(B)$ and
$\bp\sigmaqh{1}(B)=\sigmaqh{1}(B)$ with Sch{\"o}ning's BP-operator.
It is easy to see that this yields the desired claim.  Of
particular interest is to show that
$\co\up^B\nsubseteq\sigmaqh{1}(B)$. This is done by cultivating a
lower bound technique of a certain type of a real-valued
circuit. Since the acceptance probability of an oracle QTM computation
can be expressed by a multilinear polynomial of small degree
\cite{BBCMW98}, we can convert a $\sigmaqh{1,m}(A)$-computation into a
family of circuits $C$ of depth-$2$ (that work on real numbers) such
that (i) the top gate of $C$ is a $MAX$-gate of fanin at most
$m2^{n2^{n+1}}$ and (ii) all bottom gates of $C$ are polynomial-gates
of degree at most $n$ with fanin $2^n$, where a $MAX$-gate is a gate
that takes real numbers as its inputs and outputs their maximal value
and a {\em polynomial-gate} of degree $k$ refers to a gate
that computes a multilinear polynomial of degree exactly $k$. The {\em
fanin} of such a polynomial gate is the number of variables actually
appearing in its underlying polynomial. The existence of $B$ comes
from the fact that such a family of circuits cannot approximate to
within $1/3$ any Boolean function of large block sensitivity (e.g., a
$\co\up^B$-computation). 
The oracle separation at higher levels of the QP hierarchy is one of
the remaining open problems.

The third claim of Proposition \ref{prop:oracle-separation} follows
from the result $\qma^C\nsubseteq\ma^C$ \cite{Wat00} and its
generalization. The detail will appear in the complete version of this
extended abstract.

\bibliographystyle{alpha}

\end{document}